\documentclass[fleqn,10pt]{wlscirep}
\usepackage[utf8]{inputenc}
\usepackage[T1]{fontenc}
\usepackage{hyperref}
\usepackage[final]{pdfpages}

\hypersetup{
    colorlinks = false,
    linkbordercolor = {white},
}

\title{Impact of inter-city interactions on disease scaling}

\author[1]{\normalsize Nathalia A. Loureiro} %ORCID: 0009-0009-6007-1832 \email{nath.aloureiro93@gmail.com }
\author[1]{\normalsize Camilo R. Neto} %ORCID: 0000-0001-6783-6695 \email{camiloneto@usp.br}
\author[2]{\normalsize Jack Sutton} %ORCID: 0000-0002-7635-6578 \email{j.sutton@derby.ac.uk}
\author[3,4,5,6]{\normalsize Matja{\v z} Perc} %ORCID: 0000-0002-3087-541X \email{matjaz.perc@gmail.com}
\author[7,*]{\normalsize Haroldo V. Ribeiro} %ORCID: 0000-0002-9532-5195 \email{hvr@dfi.uem.br}

\affil[1]{\normalsize Complex Systems Modeling Program, School of Arts, Sciences and Humanities, University of S\~ao Paulo, S\~ao Paulo, Brazil}

\affil[2]{\normalsize College of Science and Engineering, University of Derby, Markeaton Street, Derby DE22 3AW, United Kingdom}

\affil[3]{\normalsize Faculty of Natural Sciences and Mathematics, University of Maribor, Koro{\v s}ka cesta 160, 2000 Maribor, Slovenia}
\affil[4]{\normalsize Community Healthcare Center Dr. Adolf Drolc Maribor, Vo{\v s}njakova ulica 2, 2000 Maribor, Slovenia}
\affil[5]{\normalsize Complexity Science Hub Vienna, Josefst{\"a}dterstra{\ss}e 39, 1080 Vienna, Austria}
\affil[6]{\normalsize Department of Physics, Kyung Hee University, 26 Kyungheedae-ro, Dongdaemun-gu, Seoul, Republic of Korea}

\affil[7]{\normalsize Departamento de F\'isica, Universidade Estadual de Maring\'a, Maring\'a, PR 87020-900, Brazil}

\affil[*]{\normalsize Correspondence to hvr@dfi.uem.br}

\begin{abstract}
Inter-city interactions are critical for the transmission of infectious diseases, yet their effects on the scaling of disease cases remain largely underexplored. Here, we use the commuting network as a proxy for inter-city interactions, integrating it with a general scaling framework to describe the incidence of seven infectious diseases across Brazilian cities as a function of population size and the number of commuters. Our models significantly outperform traditional urban scaling approaches, revealing that the relationship between disease cases and a combination of population and commuters varies across diseases and is influenced by both factors. Although most cities exhibit a less-than-proportional increase in disease cases with changes in population and commuters, more-than-proportional responses are also observed across all diseases. Notably, in some small and isolated cities, proportional rises in population and commuters correlate with a reduction in disease cases. These findings suggest that such towns may experience improved health outcomes and socioeconomic conditions as they grow and become more connected. However, as growth and connectivity continue, these gains diminish, eventually giving way to challenges typical of larger urban areas -- such as socioeconomic inequality and overcrowding -- that facilitate the spread of infectious diseases. Our study underscores the interconnected roles of population size and commuter dynamics in disease incidence while highlighting that changes in population size exert a greater influence on disease cases than variations in the number of commuters.
\end{abstract}

\begin{document}

\rfoot{\small\sffamily\bfseries\thepage/13}%

\flushbottom
\maketitle

\thispagestyle{empty}

\section*{Introduction}

The global population has increasingly shifted toward urban areas, with a notable milestone reached in 2007 when, for the first time, the world's urban population surpassed 50\%~\cite{WorldBankPop}. This trend has continued, with the urban population reaching 56.2\% in 2020~\cite{WorldBankPop}. Projections suggest that by 2050, 68\% of the global population will live in urban areas, with an estimated 2.5 billion additional people moving to cities, primarily in Asia and Africa~\cite{WorldBankPop, jiang2017global}. This growing urban concentration presents both opportunities and challenges. Urban settings foster intellectual and economic development through the close proximity of individuals, enabling the exchange of knowledge and services. However, this growth also exacerbates issues such as environmental pollution, rising housing costs, and congestion, while straining resources like food, energy, and water. Furthermore, urbanization has far-reaching impacts on public health, financial markets, and the global economy~\cite{WorldBankPop, jiang2017global}.

The increasing concerns about the impact of cities on these various issues, together with the growing availability of extensive datasets on urban indicators, have enabled deeper investigations into urban dynamics. This has led to the emergence of a new urban science, which conceptualizes cities as complex systems shaped by the interactions among residents and between residents and urban infrastructures~\cite{batty2013new, westscale2017, dacci2025urban}, potentially explaining the emergence of diverse urban phenomena~\cite{batty2013new, westscale2017, dacci2025urban}. Urban scaling is a prominent example of emergent behavior of city systems that has captured the attention of researchers since the 1970s~\cite{warntz1967concepts, nordbeck1971urban}, but that only gained significant recognition within the complexity science community following the works of West, Bettencourt and coauthors~\cite{kuhnert2006scaling, bettencourt2007growth, bettencourt2013origins}. The so-called urban scaling hypothesis posits that the relationship between a given urban indicator $Y$ and the population $N$ within an urban system with multiple units (such as metropolitan areas, counties, or municipalities) follows a power-law function, $Y \sim N^{\beta_N}$, where $\beta_N$ is the urban scaling exponent. Urban scaling is typically categorized into three regimes:~\cite{bettencourt2007growth} isometry ($\beta_N\approx1$ for indicators related to individual needs); superlinear allometry, or increasing returns to scale ($\beta_N>1$ for socioeconomic metrics); and sublinear allometry, or decreasing returns to scale ($\beta_N<1$ for infrastructure-related indicators). Similar to the explanation for the allometric scaling in biology~\cite{west1997general}, the emergence of this pattern is attributed to the underlying structure of urban networks (e.g., transport, supply, and social networks) that mediate the interactions among the city's parts~\cite{bettencourt2013origins}.

Empirical validation of the urban scaling hypothesis has driven numerous studies using data from multiple countries and examining a wide array of urban indicators. In the realm of disease incidence, the seminal work by Bettencourt {\it et al.} reported that HIV/AIDS cases in the USA scale superlinearly with population size~\cite{bettencourt2007growth}, a pattern also observed in Brazil by Antonio {\it et al.}~\cite{antonio2014growth}. Rocha {\it et al.}~\cite{rocha2015non} extended this analysis by investigating the scaling of various health indicators using data from Brazil, Sweden, and the USA. Their findings revealed that infectious diseases such as HIV/AIDS, chlamydia, and influenza generally scale superlinearly with population size. However, other infectious diseases, including leprosy, viral hepatitis, and dengue, exhibited isometric or even sublinear scaling. The authors suggest that the superlinear scaling observed in some infectious diseases likely reflects the increased number of contacts among residents in larger cities compared to smaller towns~\cite{schlapfer2014scaling}. In contrast, the isometric and sublinear scaling of other infectious diseases may be due to the insufficient allocation of medical resources in smaller cities. Patterson-Lomba {\it et al.}~\cite{patterson2015per} reported that sexually transmitted diseases scale superlinearly with the population of urban areas in the USA, even when controlling for socioeconomic variables. They also found that income inequality positively correlates with disease incidence, while educational level is negatively correlated. 

Expanding on these findings, Bilal {\it et al.}~\cite{bilal2021scaling} investigated the scaling of various mortality indicators across ten Latin American countries and the USA, identifying distinct scaling regimes among different regions and causes of death. Their results suggest the absence of a universal scaling law for mortality indicators encompassing both communicable and non-communicable diseases. This lack of universality was attributed to variations in urban characteristics influencing health, such as socioeconomic, environmental, healthcare, and behavioral factors, across different regions. Additionally, Patterson-Lomba and G\'omez-Lievano~\cite{patterson2023scaling} demonstrated that intrinsic features of diseases may also affect the scaling regime, with diseases that transmit less easily increasing at a faster pace with population than those that are more contagious. Studies on COVID-19 cases have also shown that larger USA cities initially experienced more pronounced growth rates in the number of cases~\cite{stier2021early}, with the number of cases and deaths in Brazilian cities displaying a sublinear scaling regime only during the first approximately one hundred days following the disease introduction, after which a superlinear scaling regime emerged~\cite{ribeiro2020city}.

The universality of urban scaling laws is often explained through models and theoretical frameworks that emphasize human interactions within cities while largely overlooking inter-city processes~\cite{ribeiro2023mathematical}. However, cities do not exist in isolation; rather, they are in continuous interaction with one another, driven by processes spanning economic, social, technological, political, and cultural exchanges~\cite{guimera2005team, alderson2010intercity}, as well as through the movement of people between cities for work, education, or better living conditions~\cite{prieto2018gravity, nelson2016economic}. Although the importance of inter-city processes in shaping urban dynamics is acknowledged, relatively few studies have explicitly examined their impact on urban scaling~\cite{altmann2020spatial, ribeiro2021association, alves2021commuting, ribeiro2023mathematical, liang2024intercity}. These inter-city interactions are particularly critical for the spread of diseases, yet there is still a knowledge gap on how such interactions may influence the number of disease cases in urban areas. 

Here we bridge this gap by investigating the effect of inter-city interactions on the association between population size and the number of cases for seven infectious diseases across Brazilian cities. To do so, we use the commuting network among cities as a proxy for inter-city interactions, combined with a general scaling framework based on the economic theory of production functions~\cite{heathfield1987introduction}, which has proven useful in studies of urban carbon dioxide emissions~\cite{ribeiro2019effects} and urban wealth~\cite{alves2021commuting}. This approach allows us to describe the number of disease cases as a function of both population size and the strength of inter-city interactions, modeled by the total number of commuters (the weighted total degree of a city in the commuting network). We show these models significantly outperform the traditional urban scaling model across the seven disease types, particularly by reducing bias in large urban areas. Additionally, we assess the impact of proportional changes in population and total number of commuters on disease cases by calculating an elasticity of scale derived from our models for individual cities. This elasticity depends on the product of population and number of commuters and predicts the existence of distinct scaling regimes, depending on whether this product exceeds specific thresholds. Overall, the majority of cities display decreasing returns to scale in relation to changes in population and the number of commuters, with the proportion of cities showing this trend ranging from 95\% to 66\%. This implied that a 1\% increase in both quantities is associated with less than a 1\% increase in disease cases for most Brazilian cities. However, increasing returns to scale are also observed for all disease types, with percentages ranging from 0.4\% to nearly a quarter of Brazilian cities. In these cities, a 1\% increase in population and commuters correlates with more than a 1\% increase in disease cases. Interestingly, we also identify a few small cities that exhibit negative elasticity of scale for certain disease types, indicating that a small proportional increase in population and commuters is associated with a decrease in the number of disease cases for these cities. We also investigate the individual effects of population and commuters on disease cases, finding that most cities exhibit a less-than-proportional response in disease cases to changes in either population or commuter numbers; however, an increase in commuters is associated with a decrease in syphilis and pertussis cases in most cities, as well as in a significant number of cities for tuberculosis and viral hepatitis. Finally, we compare the relative impact of both variables, revealing that changes in population generally affect disease cases more than proportional changes in the total number of commuters.

\section*{Results}

\begin{figure*}[ht]
\centerline{\includegraphics[width=0.65\textwidth]{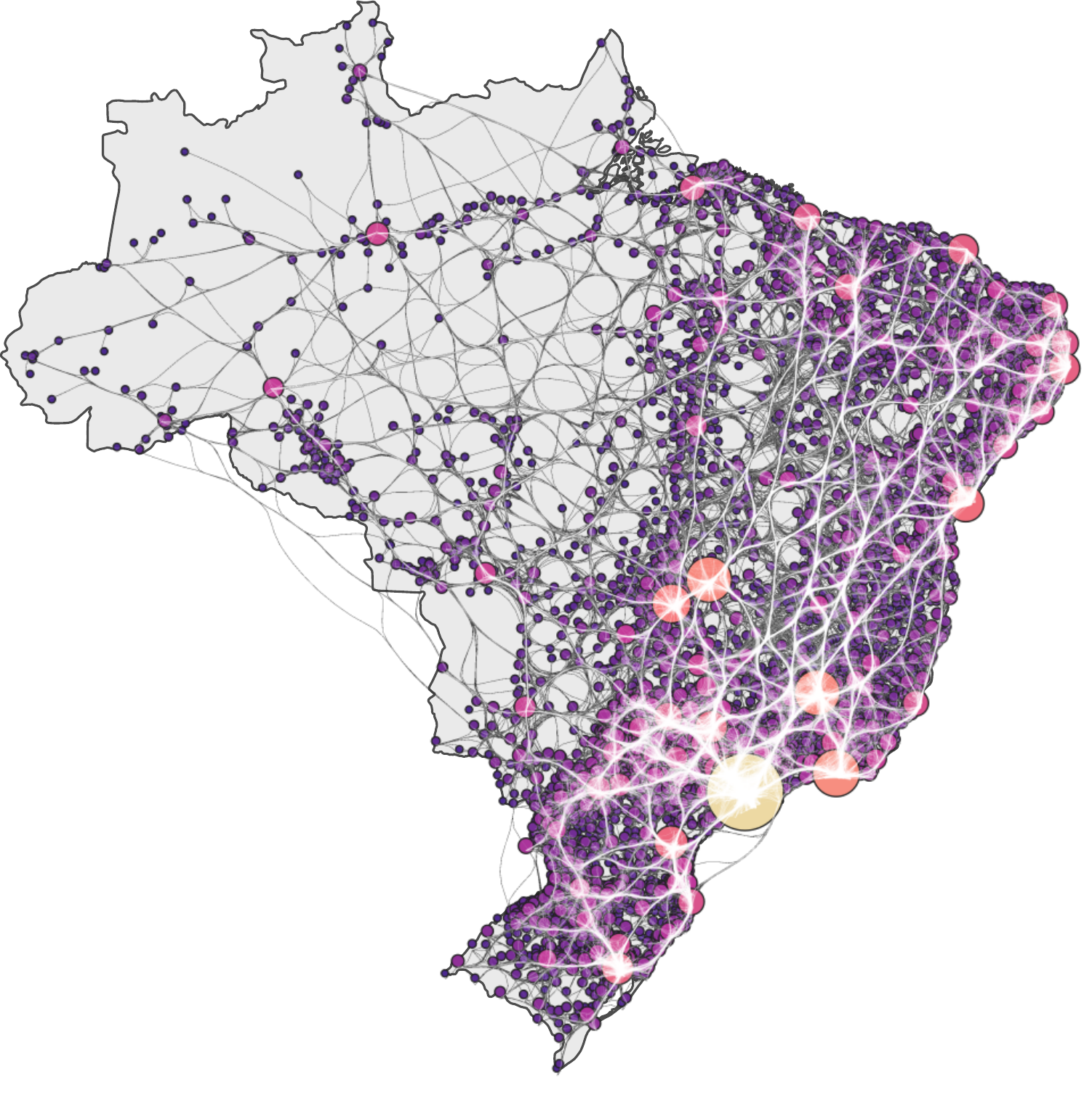}}
\caption{Commuting network among Brazilian cities. The map displays the locations of Brazilian cities, which correspond to the network nodes. Connections represent the flow of commuters between city pairs, irrespective of direction. Node sizes are proportional to the weighted total degree of the cities and are also color-coded accordingly. Edge widths indicate the number of commuters between city pairs. In this visualization, edges are grouped based on their proximity using a kernel-based edge bundling algorithm~\cite{moura20153d}. The emerging structures of this network illustrate the complexity of inter-city interactions. Figure created using Matplotlib~\cite{hunter2007matplotlib}, GeoPandas~\cite{jordahl2014geopandas}, and NetworkX~\cite{hagberg2008exploring}.
}
\label{fig:fig1}
\end{figure*}

We begin by presenting the data used in our study. The commuting network was constructed using data from the Brazilian 2010 Census, provided by the Brazilian Institute of Geography and Statistics (IBGE)~\cite{ibge2010micro}. This dataset includes information on the number of individuals who reported commuting daily from their city of residence to work in another city. We aggregate this information into a graph where nodes represent Brazilian cities, and weighted edges between city pairs indicate the total number of commuters traveling between them, irrespective of direction. The direction of commuting flow was disregarded because we use the commuting network as a proxy for inter-city interactions, and both flow directions are likely to affect disease propagation equally. Furthermore, previous research has demonstrated a strong linear correlation between the numbers of incoming and outgoing commuters in both Brazil and the USA~\cite{alves2021commuting}, indicating that considering these numbers separately does not significantly improve the models employed in our study. Figure~\ref{fig:fig1} illustrates this commuting network in which nodes are Brazilian cities and weighted edges among them indicate the total number of commuters traveling between city pairs. This representation reveals the overall complexity of inter-city interactions, which is completely ignored when cities are considered isolated entities. From this network, we evaluate the weighted total degree $S$ of each city, corresponding to the total number of incoming and outgoing commuters in a given city. This quantity indicates the centrality of cities in the commuting networks and can be further related to the overall strength of the interactions a city has with all its neighboring cities. Additionally, we obtain the population $N$ of Brazilian cities from the Brazilian 2010 Census, which is also provided by the Brazilian Institute of Geography and Statistics (IBGE)~\cite{ibge2010micro}. Finally, we collect the reported cases $Y$ of seven infectious diseases -- HIV/AIDS, influenza, pertussis, syphilis, tuberculosis, and viral hepatitis -- across Brazilian cities in 2010 from the Department of Data Processing of Brazil's Public Healthcare System (DATASUS)~\cite{datasus}. These diseases were selected based on data availability and to focus our investigation on infectious diseases transmitted directly from person to person.

Using these data, we compare the predictive power of the urban scaling model, which posits that the number of disease cases $Y$ is a function of population $N$, with a generalized framework inspired by the economic theory of
production functions~\cite{heathfield1987introduction}. In this framework, we model the number of disease cases $Y$ as the output of a two-term production function involving both the population $N$ and the total number of commuters $S$. Specifically, we consider the standard urban scaling model 
\begin{equation}\label{eq:urban_scaling}
    Y \sim N^{\beta_N} \quad \text{or its linearized form} \quad \log Y \sim \beta_N \log N \,,
\end{equation}
where $\beta_N$ is the urban scaling exponent, in comparison with two forms of the production functions borrowed from the theory of production functions~\cite{heathfield1987introduction} that have already been previously applied in urban studies~\cite{ribeiro2019effects, alves2021commuting}. The first is the Cobb-Douglas model~\cite{cobb1928theory}
\begin{equation}\label{eq:cobb_douglas_scaling}
Y \sim N^{\beta_N} S^{\beta_S } \quad \text{or its linearized form} \quad \log Y \sim \beta_N \log N + \beta_S \log S\,,
\end{equation}
where $\beta_N$ and $\beta_S$ are parameters analogous to the power-law exponent of urban scaling. The Cobb-Douglas model provides a straightforward generalization of urban scaling (which is recovered by setting $\beta_{S}=0$) while explicitly accounting for inter-city interactions mediated by the commuting network. This model exhibits a scale-invariant elasticity $\varepsilon= \beta_{N}+ \beta_{S}$, indicating that proportional changes in disease cases, in response to proportional changes in population and commuters, are independent of $N$ and $S$. Similar to standard scaling models, doubling the population and commuters can lead to less than doubling ($\beta_{N}+ \beta_{S}<1$), more than doubling ($\beta_{N}+ \beta_{S}>1$), or exactly doubling ($\beta_{N}+ \beta_{S}=1$) the number of disease cases. 

The second generalized model considered is the transcendental logarithmic (translog) production function~\cite{christensen1973transcendental, heathfield1987introduction}
\begin{equation}\label{eq:translog_scaling}
\log{Y} \sim \beta_N \log N + \beta_S \log S + \beta_C \log N\, \log S\,,
\end{equation}
where $\beta_N$, $\beta_S$ and $\beta_C$ are model parameters. The translog model extends the Cobb-Douglas model by incorporating an interaction term between population and number of commuters, modulated by the parameter $\beta_C$ (the Cobb-Douglas model is recovered when $\beta_C = 0$). This interaction term enhances the model's flexibility in describing the number of disease cases as a function of population and total number of commuters. Unlike the urban scaling and Cobb-Douglas models, the translog function does not exhibit a scale-invariant elasticity of scale. Instead, its elasticity is given by~\cite{heathfield1987introduction}
\begin{equation}\label{eq:translog_elasticity}
 \varepsilon = \beta_N + \beta_S + \beta_C \log N S\,,   
\end{equation}
indicating that variations in the number of disease cases associated with proportional changes in population and number of commuters depend on the initial values of $N$ and $S$. The translog elasticity varies from city to city, serving as a local measure of how the number of disease cases is expected to respond to changes in population and number of commuters. Thus, for fixed values of the parameters $\beta_N$, $\beta_S$, and $\beta_C$, a city can exhibit decreasing ($\varepsilon<1$), increasing ($\varepsilon>1$), or constant ($\varepsilon=1$) returns to scale, depending on whether the product of its population and number of commuters, $\Omega = N S$, is respectively smaller, larger, or equal to the critical value $\Omega^* = 10^{{(1-\beta_N-\beta_S)}/{\beta_C}}$. Interestingly, this model also allows the possibility of negative elasticity when $\Omega<\tilde{\Omega}^*$, with $\tilde{\Omega}^*=10^{{{(-\beta_N-\beta_S)}}/{\beta_C}}$. It is also informative to rewrite the translog model (Eq.~\ref{eq:translog_scaling}) by factoring either $\log N$ and $\log S$, as follows:
\begin{equation}
\log{Y} \sim (\beta_N + \beta_C\log S) \log N + \beta_S \log S \quad \text{and} \quad \log{Y} \sim \beta_N \log N + (\beta_S +\beta_C \log N) \log S\,,
\end{equation}
where the terms in brackets multiplying $\log N$ in the first expression ($\Delta_N = \beta_N + \beta_C \log S$) and $\log S$ in the second expression ($\Delta_S = \beta_S + \beta_C \log N$) correspond to marginal products of population and total number of commuters; they represent the expected change in disease cases resulting from an infinitesimal change in population and commuters. These alternative formulations of the translog model, along with the non-constant behavior of the marginal products, highlight that proportional changes in population or number of commuters lead to proportional changes in disease cases that depend on the initial values of $N$ and $S$. The translog model also allows an increase in population $N$ to be associated with a reduction in the number of disease cases $Y$ when the number of commuters $S$ is below the threshold $S^*=10^{-\beta_N/\beta_C}$. Similarly, an increase in the number of commuters can also be associated with a reduction in the number of disease cases $Y$ when the population $N$ is below the threshold $N^*=10^{-\beta_S/\beta_C}$. The threshold values $\Omega^*$, $\tilde{\Omega}^*$, $S^*$, and $N^*$ all become irrelevant when the translog parameters are positive. As we shall verify, $\beta_C$ is positive for all diseases, while $\beta_S$ is negative for all disease types, and $\beta_N$ can be either positive or negative depending on the disease type. Thus, we shall find intriguing transitions in the scaling behavior of disease cases depending on the interplay between population size and commuter numbers. 

We fit the urban scaling (Eq.~\ref{eq:urban_scaling}), Cobb-Douglas (Eq.~\ref{eq:cobb_douglas_scaling}), and translog (Eq.~\ref{eq:translog_scaling}) models to each of the seven disease types. For the urban scaling model, we estimate the value of $\beta_N$ using the standard least-squares method applied to the relationship between $\log Y$ and $\log N$ (see Supplementary Figure~S1 for visualizations of the adjusted scaling laws). In contrast, estimating the parameters of the Cobb-Douglas and translog models using ordinary least-squares is not recommended due to the strong correlations between $\log N$ and $\log S$, as well as between these terms and their product in the case of the translog model. This effect, known as multicollinearity~\cite{hastie2009elements}, occurs when two or more predictor variables in a regression model are highly correlated, leading to unstable parameter estimates and inflated standard errors. This instability arises because the ordinary least-squares method relies on the inversion of the Gram matrix $G$, defined as the product of the transpose of the regressor matrix and the regressor matrix. In the presence of strong correlations among predictors, this matrix becomes nearly singular, making its inversion highly sensitive to small perturbations in the data. To address this issue, we use the ridge regression approach~\cite{hoerl1970ridge} to estimate the parameters of the Cobb-Douglas and translog models. Ridge regression mitigates the effects of multicollinearity by adding a constant $\lambda$ to the diagonal elements of $G$, stabilizing the inversion of $G$, and reducing the sensitivity of the regression coefficients to multicollinearity. This modification is equivalent to identifying the best-fitting parameters by minimizing the residual sum of squares with an added regularization term proportional to the sum of squares of the model parameters, where $\lambda$ is the proportionally constant. Thus, in addition to the model parameters, the hyperparameter $\lambda$ needs to be estimated from the data. Following standard practice, we estimate this hyperparameter by minimizing the mean squared error using a leave-one-out cross-validation strategy. Furthermore, to ensure uniform regularization across all independent variables, we standardized the values of $\log N$ and $\log S$ before determining the optimal hyperparameter value (see Refs.~\cite{ribeiro2019effects, alves2021commuting, ribeiro2025urban} for more details). For the numerical implementation of this approach, we rely on the Python module scikit-learn~\cite{pedregosa2011scikitlearn}.

\begin{figure*}[!t]
\centerline{\includegraphics[width=0.90\textwidth]{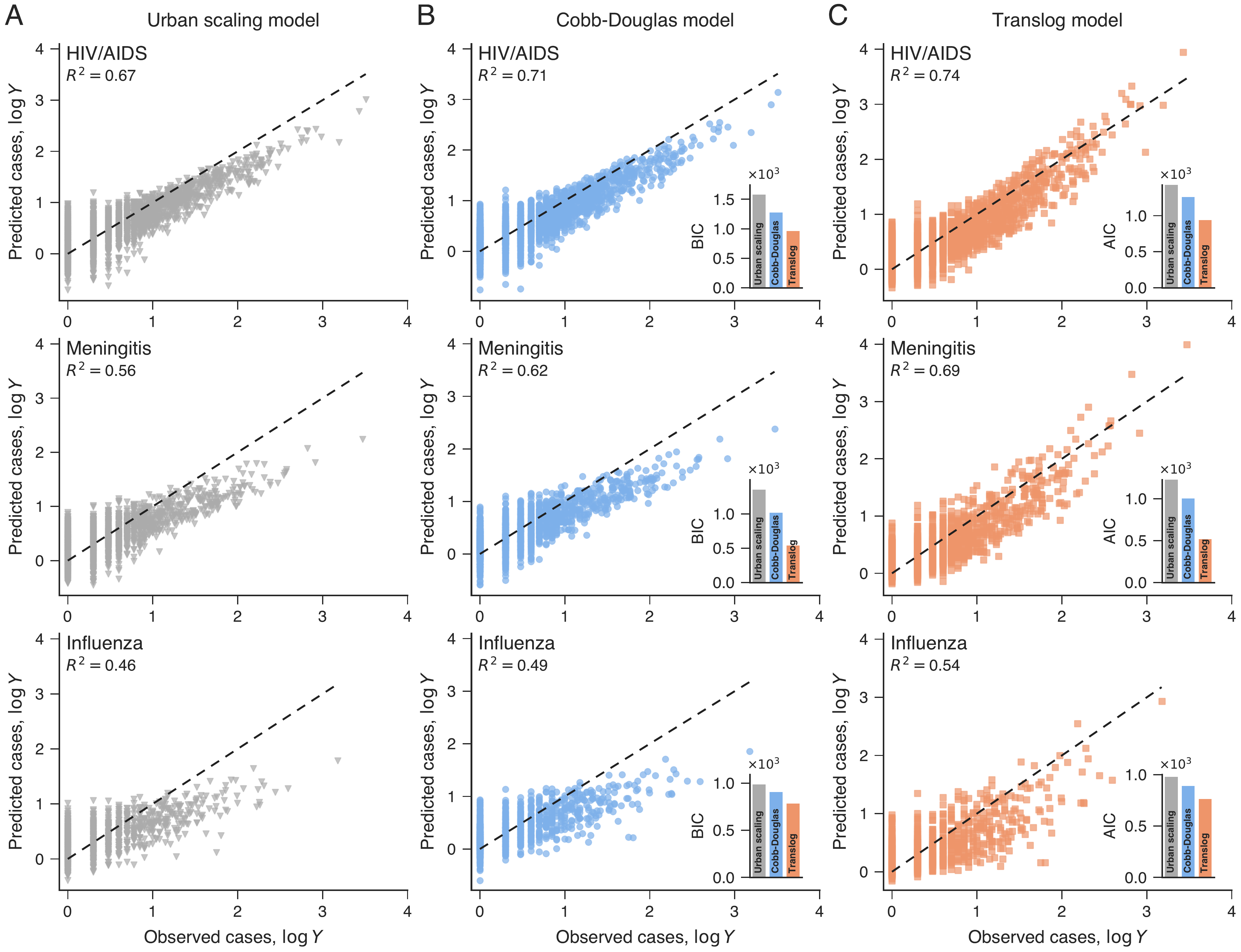}}
\caption{Urban scaling predictions of disease cases compared with the enhanced descriptions of the Cobb-Douglas and translog models. (A) Relationship between the predictions from the urban scaling model (Eq.~\ref{eq:urban_scaling}) and the observed cases of HIV/AIDS, meningitis, and influenza. (B) Improved predictions considering the number of commuters in the Cobb-Douglas model (Eq.~\ref{eq:cobb_douglas_scaling}). (C) Improved predictions considering the number of commuters in the translog model (Eq.~\ref{eq:translog_scaling}). Disease cases are expressed on a base-10 logarithmic scale, and the dashed lines represent the identity function. Insets in panels (B) and (C) compare the Bayesian information criterion (BIC) and Akaike information criterion (AIC) calculated for each model. The translog models yield the lowest BIC/AIC values and the highest coefficients of determination ($R^2$, shown within each plot), confirming a superior fit of the translog model.}
\label{fig:fig2}
\end{figure*}

Figure~\ref{fig:fig2} compares the performance of the three models in predicting the number of cases of HIV/AIDS, meningitis, and influenza. A simple visual inspection reveals that urban scaling models provide the poorest predictions (Fig.~\ref{fig:fig2}A), significantly underestimating the number of disease cases in large cities. The Cobb-Douglas models (Fig.~\ref{fig:fig2}B) improve the predictions by slightly reducing this bias. However, it is the translog models (Fig.~\ref{fig:fig2}C) that offer the most accurate predictions, markedly reducing the underestimation of disease cases in large cities. This visual assessment is corroborated by the coefficients of determination ($R^2$), shown in the figures, which attain the highest values for the translog models. Similar conclusions are drawn from the Bayesian information criterion (BIC) and Akaike information criterion (AIC)~\cite{burnham2003model}, which account for the varying number of parameters among the urban scaling, Cobb-Douglas (one more parameter than the urban scaling), and translog models (one more parameter than the Cobb-Douglas). Specifically, the insets of Figures~\ref{fig:fig2}B and \ref{fig:fig2}C compare the BIC and AIC values across the three models, demonstrating that the translog model yields the minimum BIC and AIC values, indicating not only the best predictions but also the most parsimonious fit for our data.

We have illustrated the superior predictive performance of the translog model for HIV/AIDS, meningitis, and influenza, but the same holds for all disease types in our study (see Supplementary Figure~S2). Indeed, Figure~\ref{fig:fig3} demonstrates that translog models display the lowest values of BIC and AIC as well as the highest values of $R^2$ across the seven disease types. The Cobb-Douglas model offers the second-best description for five of the diseases, but its improvement is marginal for tuberculosis and viral hepatitis. These results indicate that the interaction term between population and number of commuters in the translog model is the main factor driving the improved predictions compared to the urban scaling model.

\begin{figure*}[!ht]
\centerline{\includegraphics[width=0.77\textwidth]{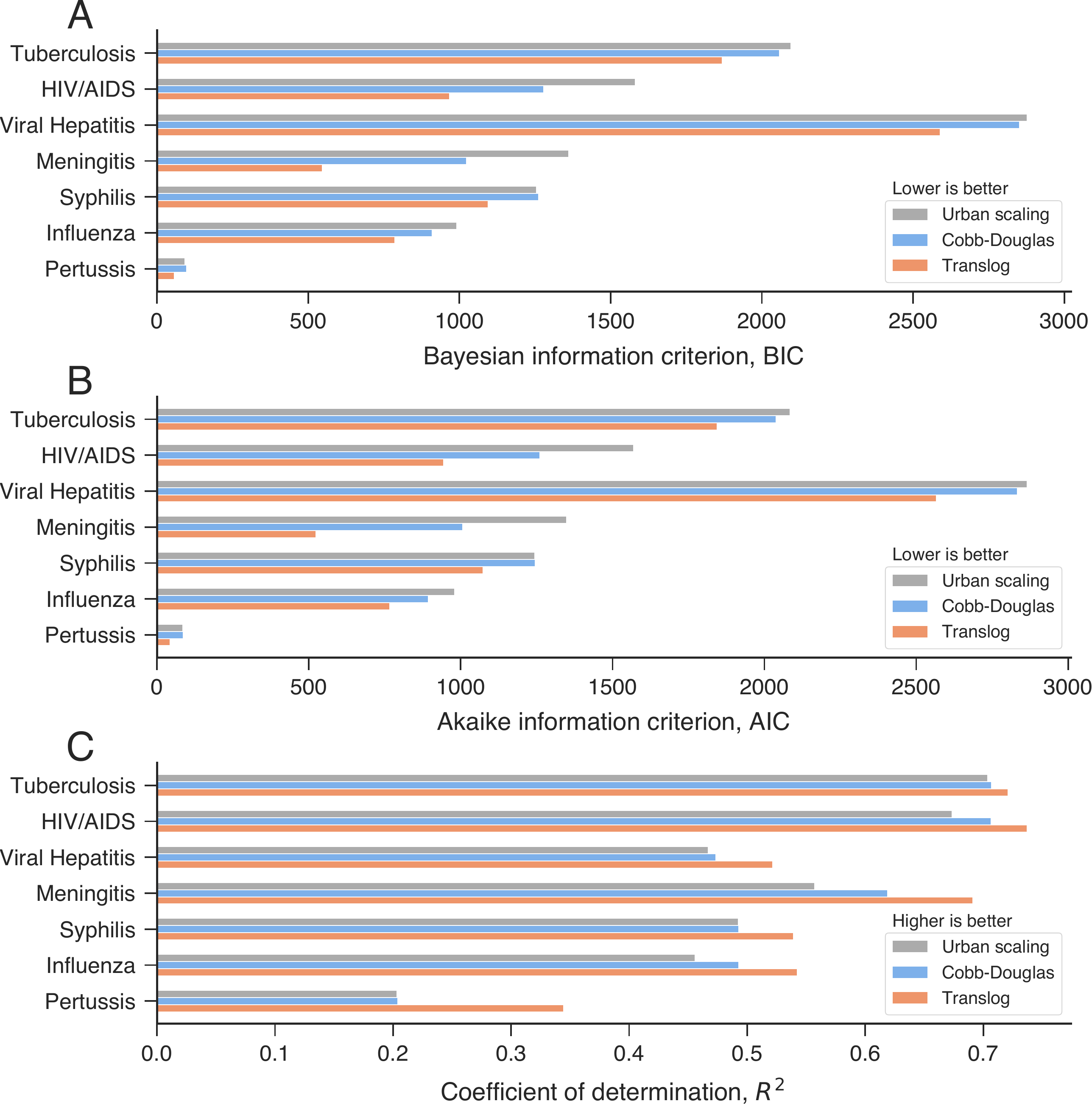}}
\caption{Comparison of the goodness of fit among the urban scaling, Cobb-Douglas, and translog models. Values of the (A) Bayesian information criterion (BIC), (B) Akaike information criterion (AIC), and (C) coefficient of determination ($R^2$) calculated for the three models across the seven disease types. The gray bars represent values for the urban scaling model, blue bars for the Cobb-Douglas model, and red bars for the translog model. The translog model provides the best (and most parsimonious) fit for all disease types according to the three model selection criteria.}
\label{fig:fig3}
\end{figure*}

\begin{figure*}[ht]
\centerline{\includegraphics[width=0.95\textwidth]{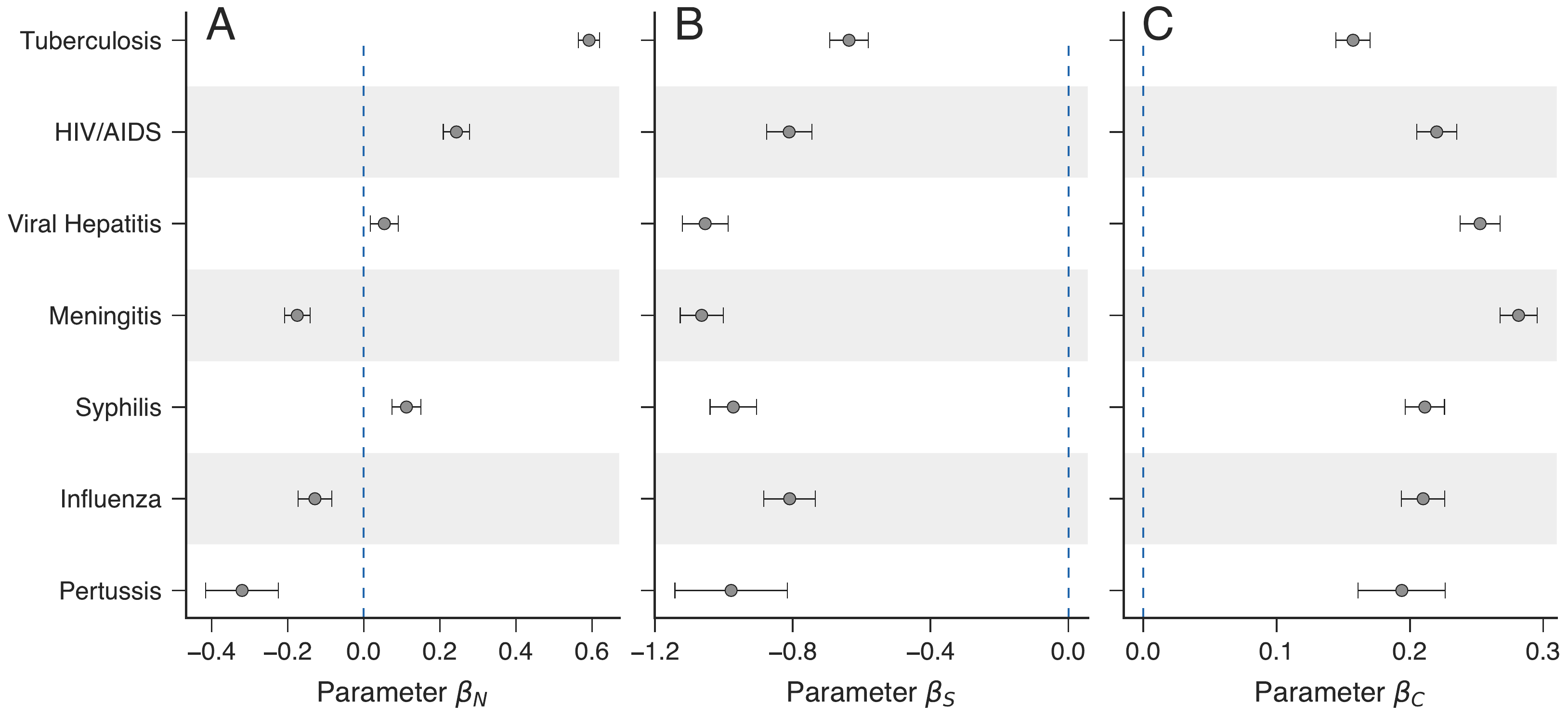}}
\caption{Parameters of the translog model estimated for each disease type. (A) Parameter $\beta_N$ quantifying the association between population size and disease cases. (B) Parameter $\beta_S$ quantifying the association between the number of commuters and disease cases. (C) Parameter $\beta_S$ quantifying the combined effect of population size and number of commuters on disease cases. In all panels, the error bars represent the standard errors of the parameter estimates. All parameters are statistically significantly different from zero.}
\label{fig:fig4}
\end{figure*}

As the translog model provides the most parsimonious and accurate description of our data, we focus on interpreting its adjusted behavior for each disease type. Figure~\ref{fig:fig4} presents the estimated values of $\beta_N$, $\beta_S$, and $\beta_C$ across all disease types. We observe that $\beta_S$ is negative and $\beta_C$ is positive for all disease types. In contrast, $\beta_N$ is positive for tuberculosis, HIV/AIDS, viral hepatitis, and syphilis, while it is negative for meningitis, influenza, and pertussis. However, analyzing these parameters individually is insufficient to fully understand the effects of changes in population and commuters on disease incidence due to the coupled nature of the translog model.

\begin{figure*}[ht]
\centerline{\includegraphics[width=1\textwidth]{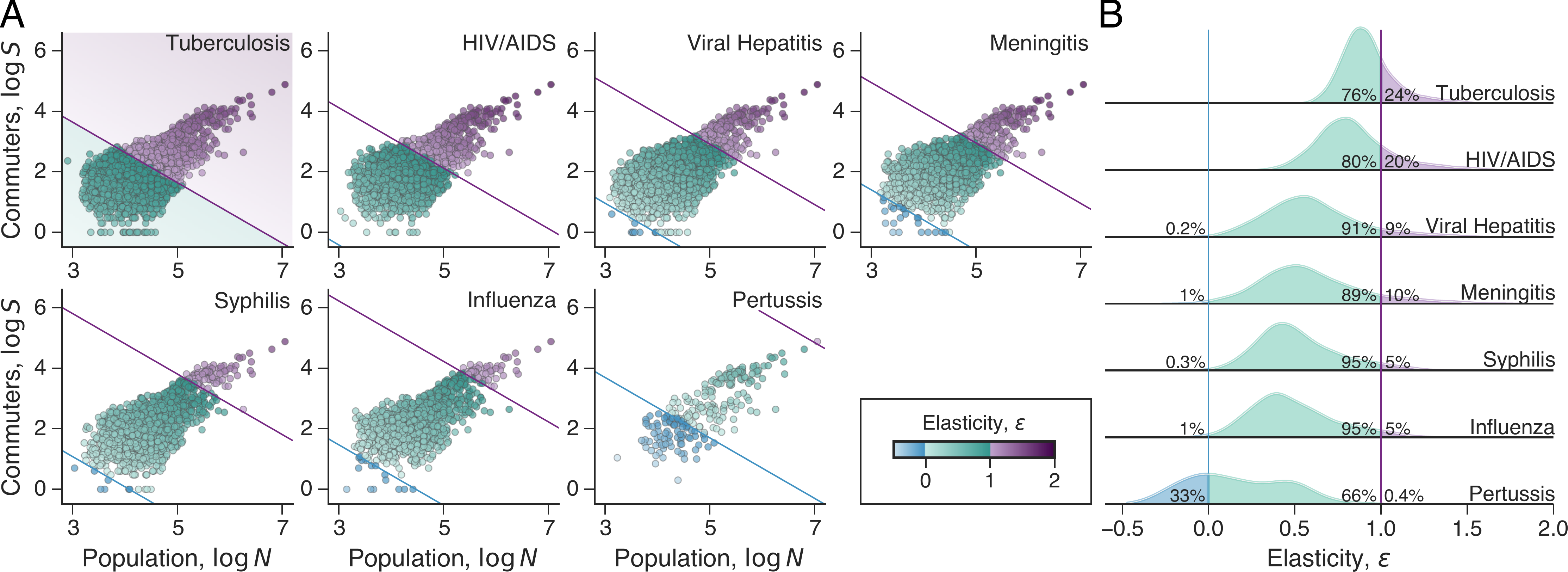}}
\caption{Elasticity of scale for each city and transitions in the scaling regime. (A) Dependence of the elasticity of scale $\varepsilon$ on population and commuter numbers for every Brazilian city reporting cases of tuberculosis, HIV/AIDS, viral hepatitis, meningitis, syphilis, influenza, and pertussis. The color-coded markers and background colors in each panel correspond to the values of $\varepsilon$, while the blue and purple continuous lines indicate the transitions from negative to sublinear and from sublinear to superlinear scaling regimes, respectively. (B) Probability distribution of the elasticity of scale estimated for each disease type. The three colors represent the fractions of cities within negative (blue), sublinear (green), and superlinear (purple) scaling regimes, with vertical lines indicating the transitions between them.}
\label{fig:fig5}
\end{figure*}

To address this, we first calculate the elasticity of scale $\varepsilon$ (as defined by Eq.~\ref{eq:translog_elasticity}) for each disease type and city in our dataset. Since this elasticity also depends on both the population $N$ and the total number of commuters $S$, Figure~\ref{fig:fig5}A depicts $\varepsilon$ using a color gradient in a scatter plot of $\log S$ versus $\log N$. For cities with $\varepsilon>1$ (purple hues), a 1\% increase in both population and commuters is associated with more than a 1\% increase in disease cases. Conversely, for cities with $0<\varepsilon<1$ (green hues), a less than 1\% increase in disease cases is expected for the same 1\% increase in both population and commuters. Lastly, for cities where $\varepsilon<0$ (blue hues), an increase in both population and commuters correlates with a reduction in disease cases. In this figure, continuous lines and varying background colors delineate the distinct scaling regimes, which are obtained by solving Eq.~\ref{eq:translog_elasticity} for each threshold. We observe the three transitions in the scaling regimes for viral hepatitis, meningitis, syphilis, and influenza. In contrast, tuberculosis and HIV/AIDS only present transitions from sublinear to superlinear scaling, while pertussis shows only a transition from negative to sublinear scaling, except for the one sole city with $\varepsilon \approx 1$. 

We also evaluate the probability distributions of $\varepsilon$ for each disease. Figure~\ref{fig:fig5}B shows these distributions, indicating that decreasing returns to scale is the most common regime across the seven disease types. The percentage of cities exhibiting sublinear scaling ranges from 66\% for pertussis to 95\% for syphilis and influenza. Increasing returns to scale represent the second most common regime, with the percentage of cities ranging from 24\% for tuberculosis to 5\% for syphilis and influenza. Negative scaling is the rarest behavior; although present in a few cities for viral hepatitis, meningitis, syphilis, and influenza, it is only significantly represented among cities in the cases of pertussis (33\%). To better illustrate the implications of these distinct regimes, consider the case of HIV/AIDS. A 1\% increase in population and commuters translates into an approximately 2\% rise in HIV/AIDS cases in S\~ao Paulo and Rio de Janeiro (the two largest metropolises in Brazil), while the same change is associated with a 0.86\% rise in HIV/AIDS cases in Parintins (a city with 100,000 inhabitants in the far east of the Amazonas state) and an exact 1\% increase in Guarapuava (a city with 180,000 inhabitants in the center of the state of Paran\'a). Now, considering pertussis, a 1\% increase in population and commuters results in the same proportional increase of disease cases in S\~ao Paulo, while the same change is expected to reduce disease cases by 0.18\% in Santana do Araguaia (a city with 75,000 inhabitants in the southernmost part of the state of Par\'a). In addition to these distributions, investigating the geographic patterns in the spatial distribution of elasticity values offers an intriguing avenue for future research. While a comprehensive analysis is beyond the scope of this study, preliminary mapping of these values (see Supplementary FigureS3) suggests the existence of correlated structures akin to those observed in the spatial distribution of disease incidence~\cite{gallos2012collective,antonio2017spatial} and other urban indicators~\cite{alves2015spatial,alves2019hidden}.

Somewhat counterintuitively, negative scaling regimes have been observed in density scaling laws, particularly in the context of housing prices~\cite{hanley2016rural, sutton2020rural}, where the price of detached housing decreases with increasing population density at high densities in England. In our study, cities exhibiting negative elasticities are predominantly observed for pertussis cases. These cities are mainly located in the inner regions of Brazil (see Supplementary Figure~S4), which are often characterized by smaller populations, lower economic development, and limited healthcare resources compared to larger urban centers. Pertussis, or whooping cough, is a highly contagious respiratory disease that primarily affects infants and young children. It is a vaccine-preventable disease, and Brazil has included the pertussis vaccine in its National Immunization Program since the 1970s, offering it free of charge through the public healthcare system. We hypothesize that as these small and isolated cities grow and become better connected, they may benefit from improved health services and socioeconomic conditions. These improvements may include more aggressive vaccination campaigns and increased awareness of disease prevention, which can more effectively reduce risky behaviors that facilitate the spread of pertussis than in smaller and more isolated populations. However, these potential benefits appear to saturate as the population and total number of commuters continue to rise. We further hypothesize that a similar mechanism may be at play in the few cities displaying negative elasticities for other diseases. Nevertheless, the specific characteristics of these diseases, such as differing transmission dynamics and latency periods, may attenuate this initial benefit, resulting in a significantly smaller number of cities in this regime.

As previously mentioned, decreasing returns to scale is the predominant response of cities to a proportional increase in both population and commuters. This regime is more common among cities of intermediate size and connectivity within the commuting network. As these cities grow and enhance their connectivity, they may experience modest improvements in healthcare and socioeconomic conditions compared to cities with negative scaling while also beginning to encounter challenges typical of larger urban areas, which contribute to increased disease transmission. The balance of these factors may yield sublinear regimes with variations across disease types. In contrast, increasing returns to scale tend to emerge in large, highly connected cities. The transition from decreasing to increasing returns to scale in disease cases is likely multifactorial, involving socioeconomic, infrastructural, and behavioral influences. Large, highly connected cities tend to feature high-density areas, more frequent social interactions, increased mobility patterns, and greater socioeconomic inequalities, all of which may contribute to environments where infectious diseases spread more efficiently. For instance, substandard housing conditions, such as overcrowded spaces and poor ventilation, are more common in large urban centers and may facilitate the airborne transmission of diseases like tuberculosis and influenza. Additionally, large cities often have higher rates of substance abuse, unsafe sexual practices, and transient relationships, which could explain the more than proportional rise in sexually transmitted infections such as HIV/AIDS and syphilis.

\begin{figure*}[ht]
\centerline{\includegraphics[width=1\textwidth]{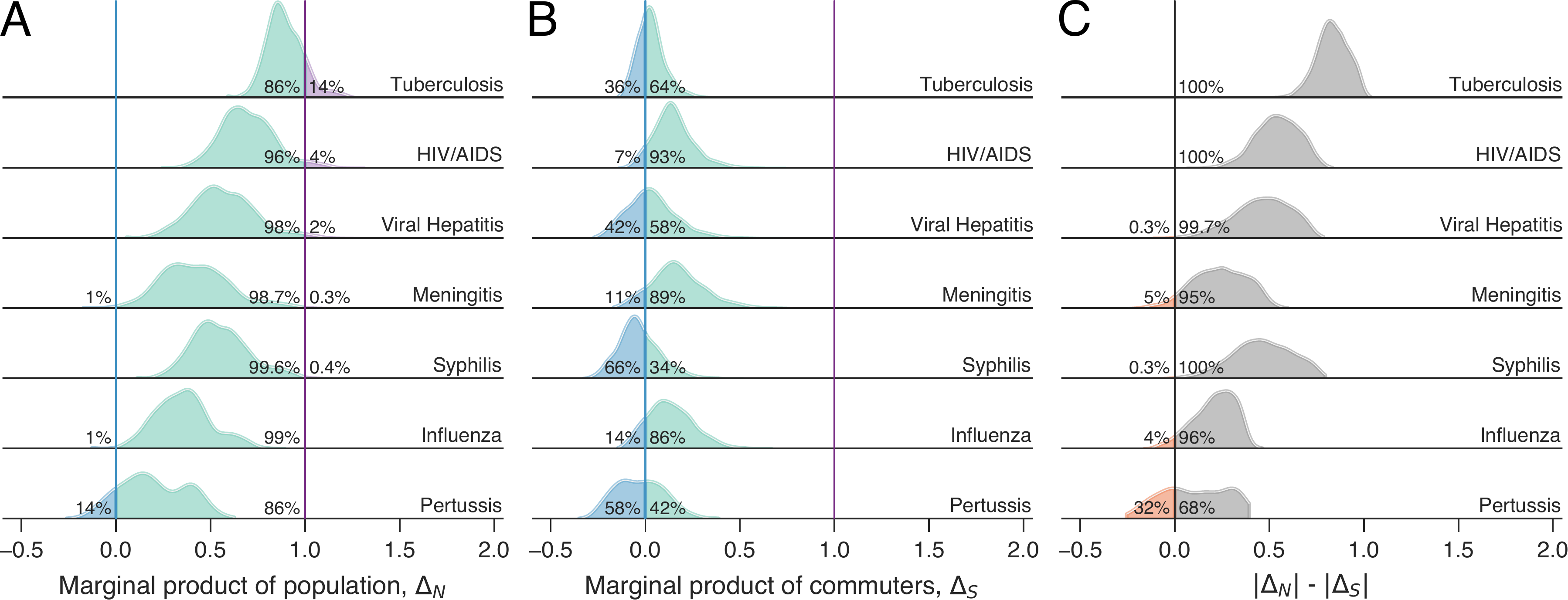}}
\caption{Marginal products of population and total number of commuters. Probability distributions of the marginal products for (A) population $\Delta_N$ and (B) commuters $\Delta_S$, calculated for all Brazilian cities reporting cases of tuberculosis, HIV/AIDS, viral hepatitis, meningitis, syphilis, influenza, and pertussis. The three colors represent the fractions of cities with negative (blue), sublinear (green), and superlinear (purple) marginal products, with vertical lines marking the transitions between them. (C) Probability distributions of the difference between the absolute values of the marginal products of population and commuters ($|\Delta_N| - |\Delta_S|$). Positive values (gray) indicate cities where changes in population have a greater impact on disease cases than changes in commuters, while negative values (red) indicate the few cities where changes in commuters have a larger effect on disease cases.}
\label{fig:fig6}
\end{figure*}

In addition to analyzing the combined effects of changes in population and commuters, we further examine the individual impacts of these variables on disease cases by evaluating the marginal products of population, $\Delta_N = \beta_N + \beta_C \log S$, and total number of commuters, $\Delta_S = \beta_S + \beta_C \log N$. As previously mentioned, $\Delta_N$ and $\Delta_S$ represent the expected changes in disease cases resulting from small proportional changes in population and commuters, respectively. Thus, a small proportional increase in population, $N \to (1+x)N$ (with $x \ll 1$), leads to a proportional increase in disease cases that depends on the parameters $\beta_N$ and $\beta_C$, as well total number of commuters, $Y \to (1+x\Delta_N)Y$. Similarly, a small proportional increase in commuters, $S \to (1+x)S$, results in a proportional increase in disease cases that depends on the parameters $\beta_S$ and $\beta_C$, and the population size, $Y \to (1+x\Delta_S)Y$. We calculate the values of $\Delta_N$ and $\Delta_S$ for each city and estimate their probability distributions for each disease type, as shown in Figures~\ref{fig:fig6}A and \ref{fig:fig6}B. Only 1\% cities exhibit negative marginal products of population for meningitis and influenza, while 14\% of cities satisfy this condition for pertussis. A small percentage of cities display values of $\Delta_N$ greater than one for syphilis, meningitis, viral hepatitis, and HIV/AIDS, while 14\% of cities satisfy this condition for tuberculosis. Thus, similar to the elasticity of scale, disease cases generally respond sublinearly to changes in population for most cities across all disease types, with percentages ranging from 86\% to 99.6\% of cities. The marginal product of commuters behaves quite differently, with no city exhibiting values of $\Delta_S$ larger than one. The most common response to changes in commuters is a less-than-proportional change in disease cases for most diseases; however, the percentages of cities in this regime range from 34\% to 93\%. A decrease in disease cases associated with an increase in commuters is the most common response of cities for syphilis and pertussis, with a significant number of cities also showing this behavior for tuberculosis and viral hepatitis. These results highlight the potential benefits of improved connectivity for cities, which may be linked to a combination of socioeconomic advantages, better access to healthcare, and more effective public health interventions in highly connected cities. The marginal products further enable us to compare the effects of changes in population and commuters on disease cases. To achieve this, we calculate the difference between the absolute values of $\Delta_N$ and $\Delta_S$ for each city and disease type, and analyze the distribution of these differences. As shown in Figure~\ref{fig:fig6}C, proportional changes in population generally have a greater impact on disease cases than proportional changes in commuters across the vast majority of cities and diseases. A notable exception is pertussis, where a significant percentage of cities exhibit a greater effect for changes in total number of commuters. Therefore, while inter-city relations proved important in improving the description of disease cases, our findings also highlight that population size alone remains a key determinant of disease incidence.

\section*{Discussion}

Our work addresses the existing gap in the literature concerning the effects of inter-city interactions on the scaling of disease cases in urban areas. By extending traditional urban scaling models to incorporate both population size and total number of commuters, we demonstrate that inter-city interactions play a critical role in shaping disease incidence across Brazilian cities. Our results indicate that models that account for the number of commuters outperform those based solely on population size, particularly in large urban centers where traditional scaling models tend to underestimate disease cases. Moreover, our generalized models reveal distinct scaling regimes for disease incidence, driven by the interplay between population size and number of commuters.

We find that the majority of cities exhibit sublinear scaling, wherein proportional increases in both population and commuters are associated with less-than-proportional increases in disease cases. However, a significant subset of cities (ranging from 5\% to almost a quarter of Brazilian cities) exhibits superlinear scaling, where a 1\% increase in population and commuters results in a more-than-proportional rise in disease cases. Interestingly, our analysis identifies a small number of cities where proportional increases in population and commuters are associated with a reduction in disease cases, particularly for pertussis. Superlinear scaling is most common in large, highly connected cities, suggesting that socioeconomic and infrastructural conditions -- such as overcrowding, poor ventilation, and inequality -- amplify disease transmission, especially for tuberculosis and HIV/AIDS, which are most often associated with this regime. Sublinear scaling predominates across all disease types, particularly in mid-sized cities with moderate connectivity. We hypothesize that this reflects a balance between improvements in healthcare and socioeconomic conditions as these cities grow. Conversely, negative scaling, mainly observed in small cities in Brazil's interior, suggests that these areas may benefit from enhanced health services and socioeconomic conditions as they develop and become more connected. This may include more effective vaccination campaigns and increased awareness of disease prevention, particularly for vaccine-preventable diseases like pertussis. Our findings further suggest that as cities grow larger and more connected, the benefits of connectivity and urbanization diminish, eventually giving rise to both sublinear and superlinear scaling regimes.

The marginal products of population and commuters further highlight the distinct roles these variables play in disease dynamics. We find that changes in population have a greater impact on disease cases than changes in number of commuters for most cities and disease types. Pertussis, however, stands out as a partial exception, where commuters exert a greater influence than population size in approximately one-third of cities, likely reflecting the unique transmission dynamics and public health responses associated with this disease. This underscores the central role of population size in shaping disease incidence, while also highlighting the nuanced contribution of inter-city interactions via commuting networks to urban disease dynamics. 

Our study highlights the potential benefits of improved connectivity between cities, particularly regarding public health outcomes. Increased connectivity is likely associated with a combination of socioeconomic advantages, better access to healthcare, and more effective public health interventions. These benefits are especially evident in smaller cities and those with moderate levels of connectivity, where enhanced commuting networks may mitigate disease incidence. However, as cities expand and connectivity intensifies, the challenges of increased disease transmission may outweigh the benefits, particularly for diseases that spread more easily through human contact. 

Our study is not without limitations. While our findings provide evidence supporting the suitability of the Cobb-Douglas and translog models in better describing the data, these models lack mechanistic explanations that directly link the holistic concept of interacting cities to their specific functional forms or to the broader theoretical framework of production functions from which they are derived. Additionally, despite the application of a regularization approach to estimate model parameters, the strong correlations between population size and the number of commuters may constrain the ability to disentangle their individual effects on disease cases. To address these limitations, future research could explore mechanistic foundations, explicitly incorporate correlations among predictors into model assumptions, and account for zero disease counts in cities. Recent advances in urban scaling --- such as generative processes based on the distribution of tokens among individuals~\cite{leitao2016scaling, gerlach2019testing, altmann2020spatial, altmann2024statistical} --- offer promising avenues for addressing these challenges.

Despite these limitations, our research shows that while population size remains a key determinant of disease incidence, inter-city interactions, modeled through commuting networks, add an essential layer of complexity to understanding urban disease dynamics. Incorporating networks into disease models not only improves predictive accuracy but also uncovers distinct scaling behaviors that vary across disease types and city sizes. Future research could also explore how other forms of inter-city interactions, such as trade and social networks, further influence disease transmission. Additionally, our findings suggest that urban planning and public health interventions should consider both population size and connectivity to optimize strategies for controlling infectious diseases in urban areas.

\section*{Acknowledgements}

The authors acknowledge the support of the Coordena\c{c}\~ao de Aperfei\c{c}oamento de Pessoal de N\'ivel Superior (CAPES), the Conselho Nacional de Desenvolvimento Cient\'ifico e Tecnol\'ogico (CNPq -- Grant 303533/2021-8), and the Slovenian Research Agency (Grants J1-2457 and P1-0403).

\section*{Author contributions statement}
N.A.L., C.R.N., J.S., M.P., and H.V.R. designed research, performed research, analyzed data, and wrote the paper.
 
\section*{Data availability}
The data used during the current study are freely available from the Brazilian Institute of Geography and Statistics (IBGE) and the Department of Data Processing of Brazil's Public Healthcare System (DATASUS), as well as from the corresponding authors on reasonable request.

\bibliography{references}

\clearpage
\includepdf[pages=1-4,pagecommand={\thispagestyle{empty}}]{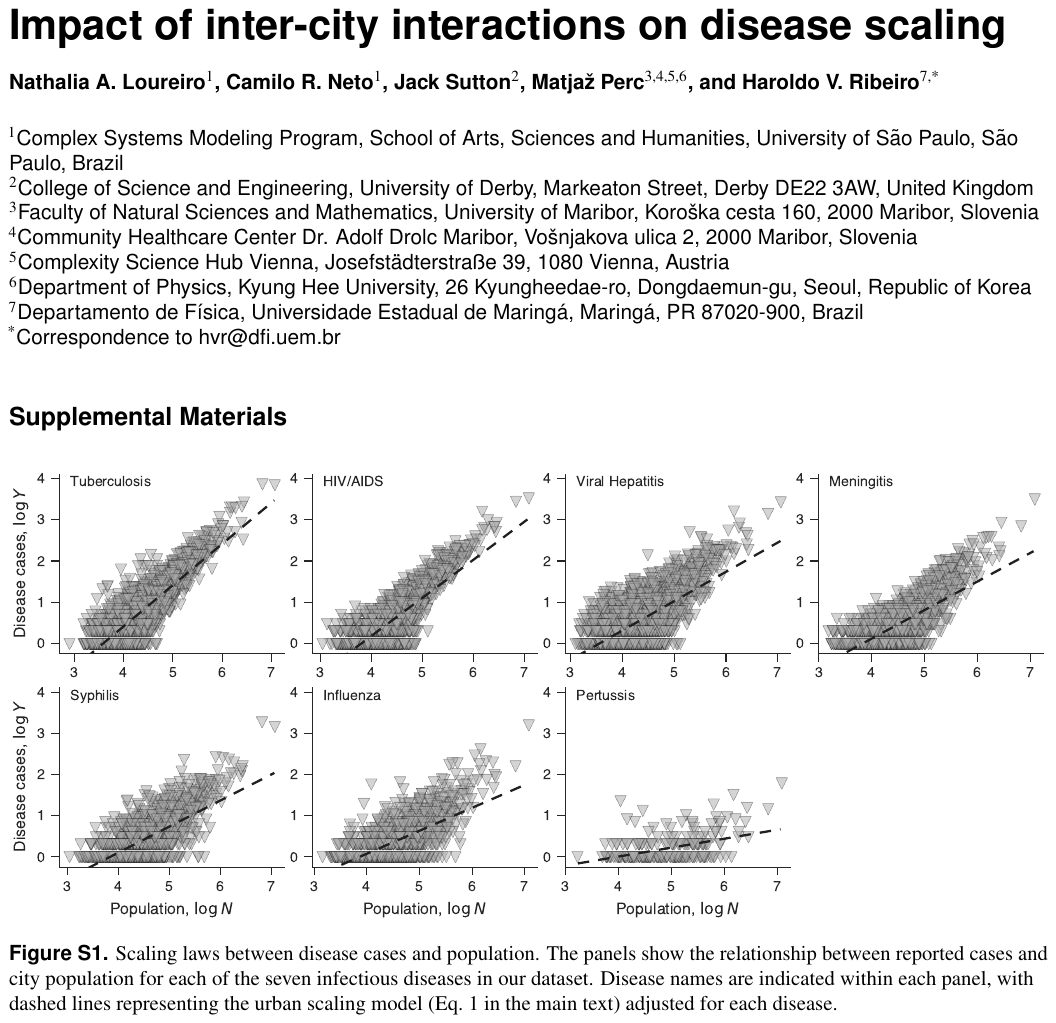}

\end{document}